# Special relativity without distant clock synchronization

Bernhard Rothenstein[1)], Stefan Popescu[2)] and George J. Spix[3)]

1) Politehnica University of Timisoara, Physics Department, Timisoara, Romania, bernhard_rothenstein@yahoo.com
2) Siemens AG, Erlangen, Germany, stefan.popescu@siemens.com
3) BSEE Illinois Institute of Technology, USA, gjspix@msn.com

***Abstract.*** *Observers at rest in two inertial reference frames are located within the propagation space of the same electromagnetic wave. Raising receiving antennas in a suitable way, these observers use the electromagnetic oscillations in the wave as an electromagnetic clock. The invariance of the wave phase ensures that the observers of the two frames detect the same phase of the oscillation when they are located at the same point in space and consequently theirs synchronization proceeds automatically. Because the Doppler Effect is free of any clock synchronization, we use the formula that accounts for this effect for deriving the basic formulas of relativistic kinematics.*

## 1. Introduction

Special relativity theory is a very flexible chapter of physics. As a consequence, we can begin its presentation with an arbitrary chosen chapter of physics and end it with another one, in an arbitrary succession, giving rise to an endless number of papers and textbooks.[1,2,3]

Karlov[4] considers that "Educators in physics can become creatures of habits like people in any field of activity. Having been taught a course in a particular way, in a sequence of steps patterned after recognized textbooks, they naturally tend to pass the same methods to successive classes of physics. From time to time, in this process, more textbooks are written which again retrace the well worn track. This perpetuation of a basic method is not necessarily something to be criticised. However, it can be a refreshing experience, when a purposeful and competent approach which deviates from the established line is encountered."

Teachers of special relativity make a net distinction between what can be derived without using the Lorentz-Einstein transformations and what can be presented as a consequence of them. Peres[5] considers that "the Lorentz transformation is the standard way to derive formulas for relativistic phenomena, such as time dilation, addition of velocities, the Doppler Effect, optical aberration, etc. Although the derivation of these formulas is straightforward, it is rather formal and not very transparent from the point of view of physics." It is surprising that he doesn't derive the Lorentz-Einstein transformation equations because the formulas he derives could lead directly to them.

While teaching the special relativity we begin by stating the principle of relativity expressed in the form of two postulates[6:]

1) The laws of physics are the same in all inertial reference frames.
2) The speed of light in free space has the same value in all inertial reference frames.

The second statement is redounding.[7,8,9] We can add to these postulates two direct consequences of the principle of relativity:
3) Distances measured perpendicular to the direction of relative motion are the same in all inertial reference frames.[10]
4) The counted number of stable objects (particles) is the same in all inertial reference frames.
5) The standard arrangement of the involved reference frames requires that we initialise the clocks of the two frames located at the corresponding origins, **when they are located at the same point in space.** Figure 1 shows a possible way in which we can perform this initialisation.

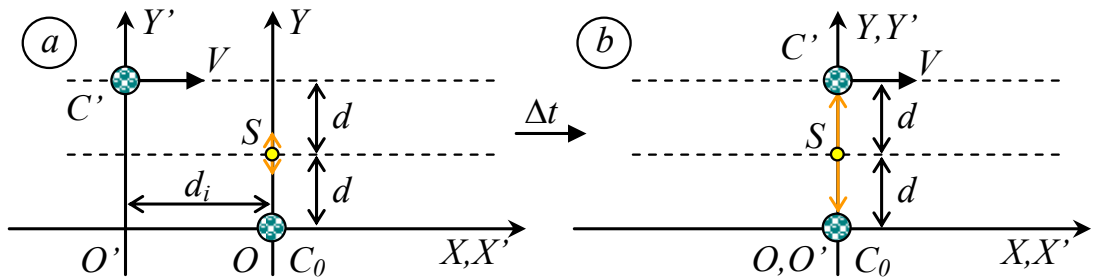

*Figure 1. Initialising two clocks situated in different reference frames to display the same time at the instant when the vertical axes overlap.*

In the first iteration let begin with the standard arrangement of the reference frames and consider a distant clock $C'$ stationary in frame K' situated at distance $2 \cdot d$ apart from the horizontal axes. In this scenario the frame K actively performs the initialisation whereas the frame K' plays only a passive role. We admit that some observers in frame K are previously synchronised and they are able to measure the horizontal speed $V$ of $C'$ and therefore to predict the instant when the clock $C'$ will reach the vertical axis OY in frame K. A light source $S$ stationary in K is pre-programmed to emit a single light pulse at instant $\Delta t = \dfrac{d}{c} = \dfrac{d_i}{V}$ before $C'$ reaches the OY axis (see figure 1a). Figure 1b displays the new situation after time $t$ elapses $\Delta t$ in frame K. The light pulses reaches simultaneously (as seen from K) both clocks and set them to display $t = t' = 0$. In the second iteration with $d \to 0$ we have $d_i \rightarrow 0$ and $C' \rightarrow C'_0$ thereby achieving the clock initialisation for the standard arrangement without any further interfering relativistic effect.

## 2. The principle of relativity in the harmonic electromagnetic wave

We consider two inertial reference frames K(XOY) and K'(X'O'Y') in relative motion. The corresponding axes of the two frames are parallel to each other, the OX(O'X') axes are overlapped, K' moving with constant velocity $V$ relative to K in the positive direction of the overlapped axes. $R_0(0,0,0)$ and $R'_0(0,0,0)$ are two observers at rest in K and K' respectively. They are located within the free propagation space of a sinusoidal electromagnetic wave (plane or spherical). These observers detect the same positive wave crest of the wave when they are located at the same point in space considered as the origin of time. Both observers raise receiving antennas perpendicular to the direction of wave propagation and parallel to the direction in which the electric component of the wave performs its sinusoidal oscillations. Under the influence of the electric vector in the wave, the free electrons in the metallic antenna perform forced oscillations. Consequently the two observers detect a harmonic potential difference between antenna ends having a period that matches the period of the electromagnetic oscillations in the wave: $\Delta\tau$ as detected from K and respectively $\Delta\tau'$ as detected from K'. The two antennas turn into clocks. The invariance of the phase[11] makes that both observers automatically detect the same oscillation phase and consequently any clock synchronization is superfluous. Counting the same number of positive wave crests $n$ passing in front of it the observer $R_0$ considers that a time $n \cdot \Delta\tau$ has elapsed, whereas the observer $R'_0$ considers that the elapsed time is $n \cdot \Delta\tau'$. By definition $\Delta\tau$ and $\Delta\tau'$ represent proper time intervals and theirs measurement doesn't involve previous synchronized clocks. The counting of the wave crests starts when the two observers being located at the same point in space detect the same wave crest. The two observers detect the same ray that propagates along a direction $\theta$ when detected from K but along a direction $\theta'$ when detected from K', both angles being measured in reference to the positive direction of the overlapped axes. After a propagation time $n \cdot \Delta\tau$ with the speed $c$, the wave crest generates in frame K the event:

$$E(x = cn\Delta\tau\cos\theta, y = cn\Delta\tau\sin\theta, t = n\Delta\tau) = E(r = cn\Delta\tau, \theta, t = n\Delta\tau),$$

while the same wave crest generates in frame K' the event:

$$E'(x' = c\,n\Delta t'\cos\theta', y' = c\,n\Delta\tau'\sin\theta', t'' = n\Delta\tau') = E'(r' = c\,n\Delta\tau', \theta', t' = n\Delta\tau')$$

expressed using Cartesian and polar coordinates as well. Let $C_0(0,0,0)$ be the antenna clock of observer $R_0$ and we consider it as representing his time measuring device similar to a wrist watch. Let $C'_0(0,0,0)$ be the antenna clock of $R'_0(0,0,0)$ being his wrist watch as well. The first problem is to find out a relationship between the proper time intervals $\Delta\tau$ and $\Delta\tau'$.

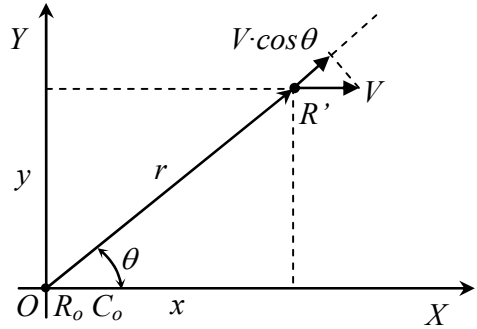

**Figure 2**. *Scenario for deriving the relationship between the changes in the readings of two distant clocks at rest relative to each other.*

Consider the experiment sketched in figure 2. A ray of the electromagnetic wave propagates along a direction $\theta$. A positive wave crest of it passes in front of $R_0$ when his clock $R_0$ reads $t_e$. After covering a distance $r$ the wave crest arrives in front of a clock $C(r,\theta)$ that reads $t_r$. The two readings are related by the obvious equation

$$t_r = t_e + \frac{r}{c} \tag{1}$$

We allow now a small change in the readings of the two clocks that leads to

$$\Delta t_r = \Delta t_e + \frac{\Delta r}{c}. \tag{2}$$

Taking into account that by definition

$$\frac{\Delta r}{\Delta t} = V\cos\theta \tag{3}$$

represents the radial velocity of an observer $R'(r,\theta)$ at rest in K' located in front of clock $C(r,\theta)$ then (2) leads to

$$\frac{\Delta t_r}{\Delta t_e} = \frac{1}{1 - \frac{V}{c}\cos\theta} \tag{4}$$

The observer $R_0$ being at rest in K measures with his clock $C_0$ a proper time interval

$$\Delta t_e = \Delta \tau. \tag{5}$$

A clock $C'(r',\theta')$ attached to observer $R'$ measures a proper time interval $\Delta\tau'$ related to $\Delta t_r$ by the time dilation formula[12]

$$\Delta\tau' = \sqrt{1-\frac{V^2}{c^2}}\Delta t_r \tag{6}$$

(4) becoming

$$\frac{\Delta\tau'}{\Delta\tau} = \frac{\sqrt{1-\frac{V^2}{c^2}}}{1-\frac{V}{c}\cos\theta}. \tag{7}$$

Considering the same experiment from the reference frame K' and following the same approach we obtain

$$\frac{\Delta\tau'}{\Delta\tau} = \frac{1+\frac{V}{c}\cos\theta'}{\sqrt{1-\frac{V^2}{c^2}}}. \tag{8}$$

We stress that $\Delta\tau'$ represents the change in the reading of clock $C'_0$ as well. The important conclusion is that (8) does not depend on the distance between the clocks measuring the proper time intervals involved. Equations (7) and (8) account for the relativistic Doppler Effect. Equating the right sides of (7) and (8) we obtain that the angles $\theta$ and $\theta'$ are related by

$$\cos\theta = \frac{\cos\theta' + \frac{V}{c}}{1+\frac{V}{c}\cos\theta'} \tag{9}$$

or

$$\sin\theta = \frac{\sqrt{1-\frac{V^2}{c^2}}\sin\theta'}{1+\frac{V}{c}\cos\theta'}. \tag{10}$$

The invariance of distances measured perpendicular to the direction of relative motion requires that

$$y = y' \tag{11}$$
$$r\sin\theta = r'\sin\theta' \tag{12}$$

or

$$r = r' \frac{1 + \frac{V}{c}\cos\theta'}{\sqrt{1 - \frac{V^2}{c^2}}} \qquad (13)$$

(10) and (13) performing the transformation of the polar coordinates of the same event. Equation (11) and

$$x = r\cos\theta = r' \frac{\cos\theta' + \frac{V}{c}}{\sqrt{1 - \frac{V^2}{c^2}}} = \frac{x' + Vt'}{\sqrt{1 - \frac{V^2}{c^2}}} \qquad (14)$$

perform the transformation of the Cartesian coordinates whereas

$$t = \frac{r}{c} = \frac{r'}{c} \frac{1 + \frac{V}{c}\cos\theta'}{\sqrt{1 - \frac{V^2}{c^2}}} = \frac{t' + \frac{V}{c^2}x'}{\sqrt{1 - \frac{V^2}{c^2}}}. \qquad (15)$$

performs the transformation of the time coordinates.

The relativistic velocities add as

$$u_x = \frac{x}{t} = \frac{u'_x + V}{1 + \frac{u'_x V}{c^2}} \qquad (16)$$

$$u_y = \frac{y}{t} = \frac{\sqrt{1 - \frac{V^2}{c^2}}}{1 + \frac{V}{c}\cos\theta'} \qquad (17)$$

Because the Doppler Effect is free of distant clock synchronization we conclude that all the relativistic equations we derived starting with the Doppler Effect are synchrony free.

### 3. Conclusions

Observers of two inertial reference frames located within the propagation space of a harmonic electromagnetic wave transform suitable oriented antennas in clocks whose period equals the period of the electromagnetic oscillations in the wave. The problem is reduced to two observers located at the origins of the two frames respectively. Because of the invariance of the wave phase both observers detect the same phase of the electromagnetic oscillations. That ensures their synchronization.

Showing that the Doppler Effect is clock synchronization free we use the formula that accounts for this effect to derive the fundamental equations of the relativistic kinematics.